\documentclass[reprint,amsmath,amssymb,aps,prl,longbibliography,lengthcheck,superscriptaddress]{revtex4}
\usepackage{graphicx}
\usepackage{dcolumn}
\usepackage{bm}
\usepackage{hyperref}
\usepackage[mathlines]{lineno}

\begin{document}
\title{Slow light in dielectric composite materials of metal nanoparticles}

\author{Kwang-Hyon Kim}
\affiliation{Max Born Institute for Nonlinear Optics and Short Pulse Spectroscopy, Max-Born-Str. 2a, Berlin D-12489, Germany}
\affiliation{Coherent Optics Department, Institute of Lasers, State Academy of Sciences, Unjong District, Pyongyang, DPR Korea}

\author{Anton Husakou}
\affiliation{Max Born Institute for Nonlinear Optics and Short Pulse Spectroscopy, Max-Born-Str. 2a, Berlin D-12489, Germany}

\author{Joachim Herrmann}
\email{jherrman@mbi-berlin.de}
\affiliation{Max Born Institute for Nonlinear Optics and Short Pulse Spectroscopy, Max-Born-Str. 2a, Berlin D-12489, Germany}

\date{\today}

\begin{abstract}We propose a method for slowing down light pulses by using composites doped
with metal nanoparticles. The underlying mechanism is related to the
saturable absorption near the plasmon resonance in a pump-probe regime,
leading to strong dispersion of the probe refractive index and significantly
reduced group velocities. By using the non-collinear scheme, it is possible
to realize the total fractional delay of 43 or larger values. This scheme
promises simple and compact slow-light on-chip devices with tunable delay
and THz bandwidth.
\begin{description}
\item[PACS numbers]
78.67.Sc, 42.25.Bs, 73.20.Mf, 78.40.Pg
\end{description}
\end{abstract}

\maketitle

\noindent Recently, slow light \cite{Khurgin(2008),Boyd(2009)} has
attracted much attention because of its fundamental significance and
promising long-reaching applications including all-optical storage,
switching and data regeneration in telecommunications, enhancement
of interferometers performance and of nonlinear optical processes.
It has been observed using different physical mechanisms such as electromagnetically
induced transparency \cite{Hau(1999),Wu(2010)}, coherent population
oscillations (CPO) \cite{Bigelow(2003),Bigelow2(2003),Cabrera-Granado(2011)},
stimulated Brillouin \cite{Okawachi(2005)} and Raman scattering \cite{Sharping(2005)},
photonic crystal waveguides \cite{Baba(2008),Hao(2010)}, fiber Bragg
gratings \cite{Mok(2006)}, double resonances \cite{Camacho1(2007)}
and others. The existing methods, however, are restricted to diverse
limitations, most notably large propagation length (such as km-scale
fibers for Brillouin scattering), small operation bandwidth (for example,
$2\sim50$ MHz in electromagnetically induced transparency), as well
as long pulse durations and others that must be circumvented for a
given application. In particular, for practical applications, miniaturized
designs of the delay line and its integration in optical networks
are requested, that can be mass-produced and will work reliably under
varying conditions. Recently progress into this direction has been
achieved by the realization of on-chip delay based on different schemes
\cite{Xia(2007),Okawachi(2006),Wu(2010)}.

In the present paper we propose a new approach for the realization
of a slow-light device which relies on composite materials doped with
metal nanoparticles (NPs). In such composites the absorption near
the plasmon resonance becomes saturated with increasing intensity
\cite{Ganeev(2003),KimKH(2010)}, because the light-induced nonlinear
change of the metal dielectric function leads to a shift of the plasmon
resonance \cite{KimKH(2010)}. The key feature of the considered system
is the retarded nonlinear response of metals in the ps time range.
Therefore in the pump-probe regime the permittivity will show large-amplitude
oscillations, if pump and probe frequencies are sufficiently close.
As a consequence, a THz-scale dip induces in a homogeneously broadened
spectral shape, and (due to Kramers-Kronig relations) a steep change
of the effective refraction index near the probe wavelength, leading
to a significant slowing down.

We consider a dielectric composite doped with small metal NPs, which
is illuminated by two waves: the strong quasi-continuous-wave pump
$E_{0}$ and the weak pulsed probe $E_{\mathrm{pr}}\left(t\right)$
with a central frequency slightly off-set from that of the pump. The
main feature in the optical response of these materials is the excitation
of plasmons in the spectral range of the plasmon resonance which is
responsible for the enhancement of the local electromagnetic field
in the vicinity of the NPs \cite{Pelton(2008)}. The total enhanced
field in the NPs can be written as $E^{\mathrm{enh}}\left(t\right)=E_{0}^{\mathrm{enh}}e^{-i\omega_{0}t}+E_{\mathrm{pr}}^{\mathrm{enh}}\left(t\right)e^{-i\omega_{pr}t}$.
Here we assume $\left\vert E_{0}^{\mathrm{enh}}\right\vert ^{2}\gg\left\vert E_{\mathrm{pr}}^{\mathrm{enh}}\right\vert ^{2}$
with a beat frequency $\Omega=\omega_{0}-\omega_{pr}$ much smaller
than $\omega_{0}$ and $\omega_{pr}$. The transient nonlinear change
of dielectric function of metal can be described as 
\begin{equation}
\Delta\varepsilon_{m}\left(t\right)=\frac{\chi_{m}^{\left(3\right)}}{\tau}\int_{-\infty}^{t}\left\vert E^{\mathrm{enh}}\left(t^{\prime}\right)\right\vert ^{2}e^{-\frac{t-t^{\prime}}{\tau}}dt^{\prime},\label{1}
\end{equation}
where $\tau$ is the electron-phonon coupling time in the range of
1-3 ps \cite{Halte(1999)} and $\chi_{m}^{\left(3\right)}=\chi_{m}^{\left(3\right)}(\omega_{0};\omega_{0},-\omega_{0},\omega_{0})$
is the inherent third-order nonlinear susceptibility of the metal
NPs at the pump wavelength. In this paper we consider picosecond pulses
and therefore the electron-electron scattering process with a response
time of few fs can be neglected. The corresponding
transient dielectric function of the metal is therefore described
by 
\begin{equation}
\begin{array}{c}
\varepsilon_{m}\left(t\right)=\varepsilon_{m0}+\chi_{m}^{\left(3\right)}\left\{ \left\vert E_{0}^{\mathrm{enh}}\right\vert ^{2}+2\mathrm{Re}\left[E_{0}^{\mathrm{enh}\ast}\tau^{-1}e^{i\Omega t}\times\right.\right.\\
\left.\left.\int_{-\infty}^{0}E_{\mathrm{pr}}^{\mathrm{enh}}\left(t+t^{\prime}\right)e^{\left(i\Omega+\frac{1}{\tau}\right)t^{\prime}}dt^{\prime}\right]\right\} 
\end{array},\label{2}
\end{equation}
where $\varepsilon_{m0}$ is the linear dielectric function of metal.
From the above equation, one can see that $\varepsilon_{m}\left(t\right)$
(and therefore the effective index) oscillates as $\cos\Omega t$.
As a result, a part of the pump beam energy is transferred to the
probe through two-beam coupling. Correspondingly, the total transmittance
of the probe is increased. This principle shows a certain analogy
with coherent population oscillations (CPO) in two level systems,
but the essential difference here is that the two beam coupling arise
by the nonlinear excitation of surface plasmons and the saturable
absorption do not arise by population inversion but by the nonlinear
shift of the plasmon resonance. The absorption dip in this mechanism
is created in nanocomposites with a \textit{homogeneous} plasmonic
absorption profile by \textit{coherent coupling} between pump and
probe, being substantially different from the slow light mechanism
based on spectral hole burning \cite{Camacho(2006)}, in which an
absorption dip originates from the selective bleaching of a homogeneous
line in the inhomogeneously broadened absorption spectrum without
any coherent beam coupling.

Separating the spectral components at $\omega_{0}$ and $\omega_{pr}$
in the electric displacement $D_{m}\left(t\right)=\varepsilon_{m}\left(t\right)E^{\mathrm{enh}}\left(t\right)$,
we find 
\begin{equation}
\begin{array}{c}
\varepsilon_{m}\left(\omega_{0}\right)=\varepsilon_{m0}+\chi_{m}^{\left(3\right)}\left\vert E_{0}^{\mathrm{enh}}\right\vert ^{2},\\
\varepsilon_{m}\left(\omega_{pr}\right)=\varepsilon_{m0}+\chi_{m}^{\left(3\right)}\left(1+\frac{1}{1+i\Omega\tau}\right)\left\vert E_{0}^{\mathrm{enh}}\right\vert ^{2}.
\end{array}\label{3}
\end{equation}
For spherical metallic NPs with diameters smaller than 10 nm, the
field enhancement factor (the ratio of enhanced field to the incident
field) is given by $x=3\varepsilon_{h}/\left(\varepsilon_{m}+2\varepsilon_{h}\right)$,
$\varepsilon_{h}$ being the permittivity of host medium. Thus we
can write the equation for the field enhancement factor of the pump
$x_{0}\left(\omega_{0}\right)$ 
\begin{equation}
x_{0}\left(\omega_{0}\right)=\frac{3\varepsilon_{h}}{\varepsilon_{m0}+2\varepsilon_{h}+\chi_{m}^{\left(3\right)}\left\vert x_{0}\left(\omega_{0}\right)E_{0}\right\vert ^{2}},\label{4}
\end{equation}
which we solve numerically \cite{KimKH(2010)}. By using the\emph{\ }standard
Maxwell-Garnett theory \cite{Maxwell Garnett(1904)} and the above
equations, we obtain the effective dielectric function for the probe.
For larger or non-spherical NPs, we apply the discrete dipole approximation
in combination with the effective medium approximation \cite{KimKH(2010)}
by using Eqs. (\ref{3}, \ref{4}). The term $1+\frac{1}{1+i\Omega\tau}$
in Eq. (\ref{3}) for $\varepsilon_{m}\left(\omega_{pr}\right)$ leads
to a narrow spectral dip in the loss coefficient and the refraction
index, the width of which is given by $\tau^{-1}$, arising from the
coherent coupling of the probe with the pump.

Let us first demonstrate the above described principal mechanism of
slow light for the simplest case of spherical metallic NPs permitting
the application of the Maxwell-Garnett formalism. Figure 1(a) shows
an example of the creation of an absorption dip and the corresponding
strong dispersion of the refractive index in a layer of silica glass
doped with Ag nanospheres smaller than 10 nm for a pump with an intensity
of 400 MW/cm$^{2}$ at the plasmon resonance wavelength of 414.7 nm.
The dielectric function of silver is taken from Ref. \cite{Palik(1985)}
and its nonlinear susceptibility $\chi_{m}^{\left(3\right)}$= $-(6.3-i1.9)\times10^{-16}$
m$^{2}$V$^{-2}$ \cite{Falcao-Filho(2005)}. The refractive index
of silica is 1.47 and the electron-phonon coupling time $\tau=1.23$
ps \cite{Halte(1999)}. The filling factor is $3\times10^{-2}$. The
transmittance for the probe strongly increases near $\Omega=0$ and
drops for higher $\Omega$. As the plasmonic dephasing time is in
the range of few femtoseconds, the width of the dip is more than 100
times narrower than the plasmon bandwidth. This narrow transparency
window in the plasmonic absorption spectrum leads to strong dispersion
and a large decrease of the group velocity near $\Omega=0$ due to
the Kramers-Kronig relation. Note that in certain parameter regions
the imaginary part of $\varepsilon_{m}(\omega_{pr})$ can be negative
and correspondingly the absorption dip change over into a gain peak.
For spherical particles this occurs if the condition $\mathrm{Im\left[\varepsilon_{m}\left(\omega_{0}\right)\right]<\left|\chi_{m}^{\left(3\right)}\left(E_{0}^{\mathrm{enh}}\right)^{2}\right|\Omega\tau/\left[1+\left(\Omega\tau\right)^{2}\right]}$
for purely real $\chi_{m}^{\left(3\right)}$, requiring stronger
pump in practice due to the non-zero imaginary part of $\chi_{m}^{\left(3\right)}$.
For spherical NPs with the above given parameters, this condition
requires a rather high intensity in the range of 10 GW/cm$^{2}.$
However, as will be shown later for gold nanorods the same effect
occurs also for non-spherical particles. Since the plasmon resonance
for nanorods can be shifted to the IR spectral range exhibiting the
much higher value of $\left|\chi_{m}^{\left(3\right)}\right|$, a
gain peak is already created for even much smaller intensities in
the range of 0.3 MW/cm$^{2}.$ 
\begin{figure}[t]
 \centerline{ \includegraphics[width=7cm]{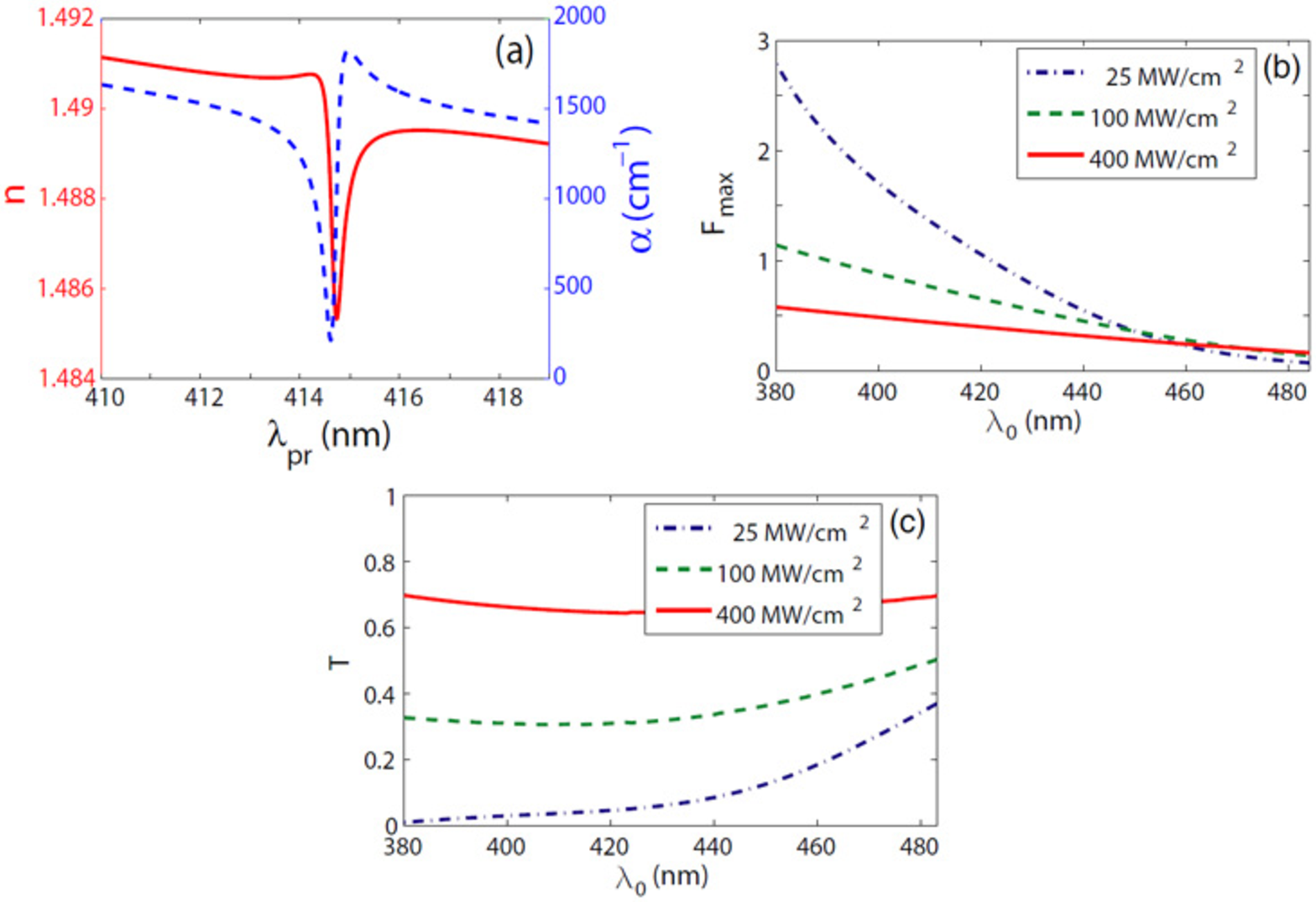}}
\caption{(Color online) Creation of absorption dip and strong dispersion of
refractive index (a) by frequency-dependent coherent energy transfer
from pump to probe, maximum fractional delay (b) and transmission
(c) as functions of the pump wavelength for different pump intensities
in a silica glass layer doped with very small Ag nanospheres. In (b)
and (c), the filling factor is 2.5$\times$10$^{-2}$, the propagation
length is 2 $\mu$m, $\tau=1.23$ ps. }
\end{figure}

Figures 1(b) and (c) show the fractional delay $F=T_{\mathrm{del}}/T_{0}$
and the transmission $T$ as functions of the pump wavelength for
different pump intensities. Here $T_{\mathrm{del}}\left(\omega_{pr}\right)=L\left(v_{g}^{-1}-c^{-1}\right)$
is the total delay, $L$ the propagation length, $c$ the light velocity
in vacuum, $v_{g}=c/n_{g}$ the group velocity, $n_{g}=d\left[n\left(\omega_{pr}\right)\omega_{pr}\right]/d\omega_{pr}$
the group index with the effective index $n\left(\omega_{pr}\right)$,
and $T_{0}$ is the pulse duration. The filling factor is $2.5\times10^{-2}$,
the propagation length is 2 $\mu$m, and the duration of the probe
pulse is 1.85 ps, which is around 1.5 times longer than the electron-phonon
coupling time $\tau$. The other parameters are the same as in Fig
1(a). One can see that the fractional delay increases with increasing
intensity while the transmission shows an opposite behavior. As can
be seen, for pump intensities below 300 MW/cm$^{2}$ the delay decreases
with increasing pump wavelength while the transmission increases.
\begin{figure}[t]
 \centerline{ \includegraphics[width=7cm]{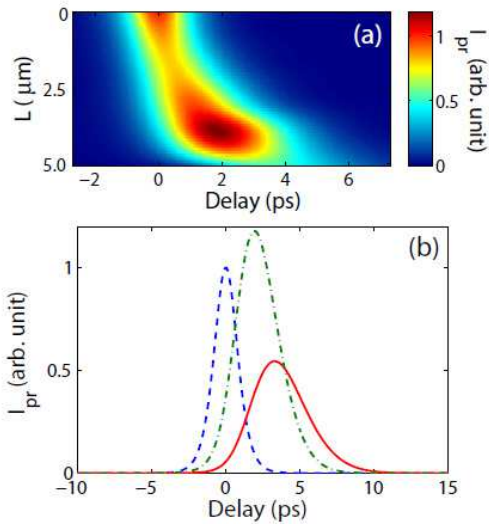}}
\caption{(Color online) Slow light in TiO$_{2}$ film doped with Au nanorods
with a diameter of 20 nm and a length of 66 nm for pump intensity
of 6 MW/cm$^{2}$ at 1550 nm. Other parameters are the same as in
Fig. 3. In (a) and (b), probe pulse evolution and optical delay are
shown, respectively. In (b), blue dotted line is the incident probe
pulse, green dash-dotted and red solid lines are probe pulses corresponding
to propagation lengths of 4 and 5 $\mu$m.}
\end{figure}

For practical applications, it is worth to study slow-light performance
at telecommunication wavelengths. The plasmon resonance can be shifted
to a large extent by tailoring the shape of the metallic NPs, in particular
nanorods are suitable to shift the plasmon resonance into the long
wavelength range. In this wavelength range, noble metals have extremely
large inherent nonlinear susceptibilities, e. g. for gold $-1.5\times10^{-11}$
m$^{2}$V$^{-2}$, which is about 4 orders larger than in the visible
range \cite{Falcao-Filho(2010)}. To take into account the decrease
of the pump intensity during the propagation due to the residual absorption
and the change in the complex amplitude of probe, here we simultaneously
solve the coupled propagation equations for the \textit{cw} pump and
the pulsed probe, using the effective dielectric function, calculated
by the method mentioned above. The time-domain probe pulse shape then
is obtained by using the Fourier transformation. 

Figure 2 displays an example of the incident and delayed pulses for
a pump intensity of 6 MW/cm$^{2}$ at 1550 nm. As a slow light material,
we take a TiO$_{2}$ film doped with Ag nanorods with a diameter of
20 nm and a length of 66 nm, exhibiting a surface plasmon resonance
at around 1540 nm. Pump and probe beams have the same propagation
direction and polarization. The pulse duration was 1.85 ps , corresponding
to $1.5\tau$. Figure 2(a) shows the evolution of probe pulse. Though
the absorption for the pump is saturated, the output pump beam is
attenuated down to 34 kW/cm$^{2}$ (not shown), which is $\sim200$
times smaller than the input pump intensity. The probe is sustained
primarily by the coherent energy transfer from the pump and then it
is attenuated by linear loss after the strong decrease of pump intensity.
The probe energy can become even larger than the incident at a certain
propagation length. The possible occurrence of a gain peak for the
probe under certain conditions was discussed above; here it arise
already for rather low intensities. This can be clearly seen from
Fig. 2(b); the peak intensity of the probe after propagation over
a length of 4 $\mu$m (green dash-dotted line) is nearly 1.2 times
larger than that of the incident pulse. For a propagation length of
5 $\mu$m, the fractional delay is about 1.79, corresponding to a
delay-bandwidth product of 2.68, with an energy transmittance (the
energy ratio of output pulse to the incident pulse) of about 0.76.
As the pulse width is broadened during the propagation in the medium,
the peak intensity was decreased slightly stronger than estimated
from the transmittance and was about 0.55. 
\begin{figure}[b]
 \centerline{ \includegraphics[width=7cm]{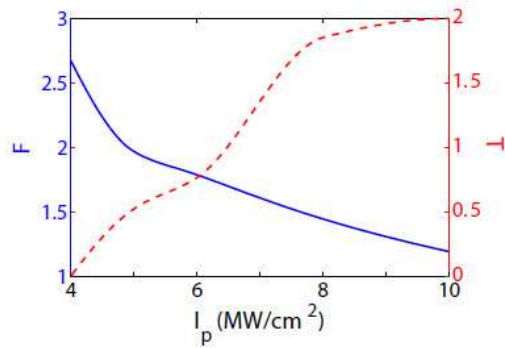}}
\caption{(Color online) Dependencies of fractional delay $F$ (blue solid line),
transmittance $T$ (green dotted line) on the pump intensity in TiO$_{2}$
film doped with Au nanorods at 1550 nm. The diameter and length of
nanorods are 20 nm and 66 nm, respectively. The propagation length
is 5 $\mu$m, the filling factor $5\times10^{-2}$, and the probe
pulse duration 1.85 ps. }
\end{figure}

The group velocity is 198 times smaller than the light velocity in
vacuum. The pulse distortion was evaluated to be around 0.76 by using
the formula \cite{Shin(2007),Bautista(2011)} 
\begin{equation}
D=\sqrt{\frac{\int_{-\infty}^{\infty}\left\vert I_{\mathrm{out}}\left(t+\triangle t\right)-I_{\mathrm{in}}\left(t\right)\right\vert dt}{\int_{-\infty}^{\infty}I_{\mathrm{out}}\left(t+\triangle t\right)dt}},\label{8}
\end{equation}
where $I_{\mathrm{in}}$ and $I_{\mathrm{out}}$ are input and output
probe pulse intensities normalized to have the same maximum, respectively,
and $\triangle t$ is the peak delay time. Because of the large inherent
nonlinear susceptibility of gold in this wavelength range, the large
fractional delay is achieved even with relatively low intensities
of few MW/cm$^{2}$.

Figure 3 presents the dependence of fractional delay $F$ and transmittance
$T$ as functions of the pump intensity in the same medium as in Fig.
2 at 1550 nm. The figure shows that the fractional delay decreases
with increasing pump intensity, while the transmittance increases.
In particular, for higher pump intensity, the transmittance for the
probe may exceed unity, indicating amplification by an energy transfer
from the strong pump. This feature is a significant advantage compared
with most of the known slow-light mechanism. Pulse distortion, in
this case, varies from 0.8 to 0.66 with increasing intensity (not
shown). Fig. 3 shows that for a given propagation length of 5 $\mu$m
a fractional delay in the range of 2 (corresponding to a delay-bandwidth
product of 3) can be obtained with a transmittance higher than 0.5. 

In our knowledge, relatively few experiments in different schemes
have measured relative pulses delays larger than these predictions.
Comparing results obtained with chip-compatible design, we refer to
Ref. \cite{Wu(2010)} with a relative delay of 0.8 or to Ref. \cite{Okawachi(2006)}
with a fractional delay of 1.33. On the other hand, considerably larger
fractional delays have been realized by using cascaded microring resonators
using photonic wire waveguides (but with relatively small probe transmission)
\cite{Xia(2007)} or by using double resonances of Cs atoms \cite{Camacho1(2007)}.
However the above given numerical examples with a fractional delay
in the range of 2 are not the physical limits of the method proposed
in the Letter but determined by the chosen design. Next we will show
that significantly larger fractional delays can be realized using
a modified arrangement.

The main limitation of the delay line in collinear configuration considered
up to now arises due to the limited propagation length caused by\emph{
}the attenuation of the pump. This can be circumvented using a transversely
pumped waveguide as shown in Fig. 4(a) where the probe pulse is guided
by the compact waveguide structure. The waveguide dispersion could
be neglected in comparison with the dispersion arising from the NPs.
The thickness of the film is taken to be 1 $\mu$m, resulting in
an attenuation of the pump intensity less than 10 \%, and, therefore,
the pump intensity can be approximated to be constant. The other parameters
are the same as in Figs. 2 and 3 except the pump intensity of 0.28
MW/cm$^{2}$. This corresponds to a pump energy of 0.2 nJ for a pump
pulse duration of 200 ps, which can be considered as quasi-continuous
wave in comparison with few ps duration of probe, and a transverse
width of the waveguide of 5 $\mu$m. Note that the polarization directions
of pump and probe have to be parallel due to the selective plasmon
excitation for non-spherical NPs. Figure 4(b) presents the evolution
of the probe pulse normalized to the peak intensity of the incident
probe pulse. In this figure, the peak intensity of the probe increases
by a factor of 3.1 after propagation over a length $L$ of 90 $\mu$m.
For detailed presentation, in Fig. 4(c) we show the normalized probe
pulses corresponding to propagation lengths up to 90 $\mu$m with
a equidistance of 6 $\mu$m\emph{.} The figure shows\emph{\ }a total
fractional delay of about 43, corresponding to a delay-bandwidth product
of 65 and a slowing down factor of around 270. The corresponding pulse
distortion remains in an acceptable range with changes from 0.60 to
0.94 with increasing propagation length. In this configuration, as
the probe can be amplified, we have no any principal limitation of
the delay time, since with increasing propagation length and
correspondingly larger pump energy the delay time can be
increased significantly only limited by the available
pump source. 
\begin{figure}[t]
 \centerline{ \includegraphics[width=7cm]{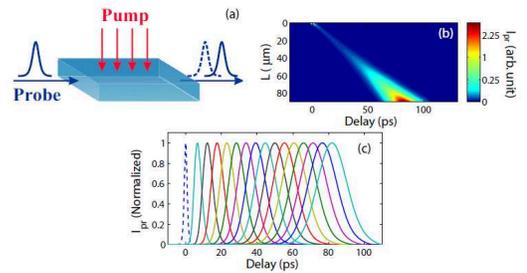}}
\caption{(Color online) Delay for probe with a duration of 1.85 ps and pump
intensity of 0.28 MW/cm$^{2}$ in TiO$_{2}$ film with thickness of
1 $\mu$m in non-collinear configuration: (a)- configuration, (b)
and (c) - evolution of probe pulse. Other parameters are the same
as in Fig. 3. In (c), blue dashed line is the incident pulse and solid
lines are the delayed pulses corresponding to propagation lengths
$L$ up to 90 $\mu$m with a equidistance of 6 $\mu$m from the left
to the right side in the order.}
\end{figure}

To conclude, we proposed and studied theoretically a slow-light mechanism
based on composites doped with metal NPs. If two pulses with a frequency
difference smaller than the electron-phonon coupling rate propagate
through such a medium, plasmon-induced oscillations of the nonlinear
permittivity arise, creating a strong dispersion of the effective
index for the picosecond probe pulses, and correspondingly a significantly
small group velocity. This scheme can be applied over a broad spectral
range from visible to infrared by tailoring the sizes and shapes of
the metal NPs. We have shown that employing these composites in a
collinear arrangement reduced group velocities of the probe with a
slow-down factor in the range of $200$, a fractional delay up to
more than 2.5 for probe pulses with a few ps duration can be realized
as at telecommunication wavelengths using gold nanorods as
well in the optical range. Avoiding pump depletion by using a non-collinear
configuration, the relative delay can be significantly increased in
a regime of probe amplification. We have shown that a total fractional
delay of 43 (delay-bandwidth product of 65) at telecommunication wavelength
can be realized in a transversely pumped waveguide geometry. Since
the probe is not attenuated but can be amplified, this value of the
fractional delay can be further increased with increasing length of
the transversely pumped waveguide. The advantage of this configuration
includes a high compactness, cost-effectiveness, broad range of applicable
spectral range with tailoring the sizes and shapes of metal NPs, wide
slow light bandwidth up to nearly THz, and large fractional delay.
This could open the perspectives for the development of simple, compact,
cheap and chip-compatible slow-light devices with large delay for
applications in optical telecommunication and other fields.

\end{document}